# Monte Carlo Modeling of Spin Injection through a Schottky Barrier and Spin Transport in a Semiconductor Quantum Well

Min Shen[a], Semion Saikin[a,b,c], Ming-C. Cheng[a,*]

[a]Department of Electrical & Computer Engineering
[a]Center for Quantum Device Technology
[b]Department of Physics
Clarkson University, Potsdam, NY 13699, USA

[c]Department of Physics, Kazan State University, Kazan 420008, Russia

[*]Contact Author: mcheng@clarkson.edu

**Abstract**

We develop a Monte Carlo model to study injection of spin-polarized electrons through a Schottky barrier from a ferromagnetic metal contact into a non-magnetic low-dimensional semiconductor structure. Both mechanisms of thermionic emission and tunneling injection are included in the model. Due to the barrier shape, the injected electrons are non-thermalized. Spin dynamics in the semiconductor heterostructure is controlled by the Rashba and Dresselhaus spin-orbit interactions and described by a single electron spin density matrix formalism. In addition to the linear term, the third order term in momentum for the Dresselhaus interaction is included.

Effect of the Schottky potential on the spin dynamics in a 2 dimensional semiconductor device channel is studied. It is found that the injected current can maintain substantial spin polarization to a length scale in the order of 1 micrometer at room temperature without external magnetic fields.

# I. Introduction

The promising application of carrier spin in modern semiconductor electronics [1-3] has initiated great interest of the scientific community in spin-polarized transport properties in semiconductor and metal-semiconductor structures. Study of spin-polarized transport can be classified into three main topics. These include spin relaxation and spin dephasing in transport, control of coherent spin dynamics and electrical spin injection/detection. Among various spintronic devices, some devices utilize spin-orbit interaction in semiconductor hetersotructures [4-9]. Spin dynamics in such devices can be controlled simply by a conventional electric gate to modulate the spin-orbit interaction mechanisms [4]. Recent experimental advances have achieved spin polarization for up to few nanoseconds at room temperature in (1,1,0) quantum wells [10]. Coherent transport of spin polarization in strained GaAs layers without external magnetic field has been observed recently for a distance longer than 60 micrometers at 5 K [11]. Gate control of the spin-obit coupling in a 2 dimensional electron gas (2DEG) has been reported for different structures [12,13]. For electrical spin injection and spin detection, two types of ferromagnetic contacts (metal [14] and semiconductor [15]) have been studied. The ferromagnetic metal/semiconductor structure offers promising device application due to the possibility to realize spin injection even at room temperature [16]. However, the experimental realization of a high-efficiency spin source remains a challenging task. The reported injection efficiency of spin polarization in such structures varies from 1%-2% up to 30%-35% for different designs and methods of measurements [16,17,18].

The fundamental issue of spin injection from a ferromagnetic metal contact into a semiconductor has been discussed in [19]. The authors performed theoretical analysis of spin injection through a perfect interface in a diffusive transport regime based on the model introduced by van Son *et al* [20]. It was found that owing to the large conductivity mismatch between two materials the efficiency of the injected spin polarization should be about 0.1% for typical device geometry. However, the approximation of a perfect interface with zero resistance is hardly applicable in many cases [21,22]. An interface layer between two materials can appreciably affect the injection efficiency. It was pointed out that a tunneling barrier at the interface between a ferromagnetic metal and a semiconductor could be a remedy to avoid the unrealistically low injection efficiency [23,24]. This theoretical prediction was verified by experimental results with different tunneling barriers [14,16,25]. The Schottky barrier is the natural barrier that appears at the ferromagnetic-metal/semiconductor interface.

In this work, we present a Monte Carlo model for spin injection through a Schottky barrier into a quantum well (QW). Such structure can be considered as a basic element for many spintronic devices [4,5,7,8]. The model is applicable beyond the drift-diffusion approximation. Both mechanisms of thermionic emission and tunneling injection are considered. The spin-polarized particles are injected from a 3D ferromagnetic metal into a III-V semiconductor QW with a Schottky barrier at the interface. The direction of injection is parallel to the QW plane. The injected spin current is mainly originated from the electrons with energy far above the Fermi level in the ferromagnetic metal, where the tunneling probability is high or the thermionic emission is possible, rather than from those near the Fermi surface. The tunneling probability through the Schottky barrier is spin dependent [26]. In the model, this dependence is approximated by the difference between densities of states for spin-up and -down electrons in the ferromagnetic contact at given energy [27]. The models for the emission and tunneling through the Schottky barrier in the Monte Carlo simulation are detailed in the next section. Transport of the spin-polarized electrons in the QW is discussed in Sec. III. The approach is based on a model developed previously [28] with an additional higher-order correction in spin-orbital coupling [29].

## II. Model of Spin Injection through a Schottky Barrier

At finite temperature, total current injected from a ferromagnetic metal to a semiconductor through the Schottky barrier can be written as [30,31]

$$j = \frac{A^* T}{k_B} \int_0^\infty T_{tp}(E) f_m(E)[1 - f_{sc}(E)] dE, \qquad (1)$$

where $k_B$ is the Boltzmann constant, $A^*$ is the Richardson constant [30], $T$ is temperature, $T_{tp}(E)$ is the tunneling probability through the barrier at the energy $E$, and $f_m$ and $f_{sc}$ are electron distribution functions in the ferromagnetic metal and semiconductor, respectively. When $E$ is above the Schottky barrier height $q\phi_B$, $T_{tp}(E) = 1$. If the current is small enough, the ferromagnetic source can be assumed in thermal equilibrium, and the energy-state occupancy obeys the Fermi-Dirac distribution,

$$f_m(E) = \frac{1}{1 + e^{(E-E_{fm})/k_B T}}, \qquad (2)$$

where $E_{fm}$ is the Fermi level in the metal. Electrons injected into the semiconductor in the depletion region experience strong electric field and may gain high kinetic energy within a short distance. Implementation of the injection and transport through the Schottky barrier in the Monte Carlo model is presented below.

For an electron in the metal contact, $E$-$E_{fm}$ is randomly partitioned into three components associated with $x$, $y$ and $z$ directions, $E - E_{fm} = E_x + E_y + E_z$, with the expectation value in each direction equal to one third of $E$-$E_{fm}$. The spin state of the electron is determined by the probabilities

$$P_\uparrow(E) = \frac{D_\uparrow(E)}{D_\uparrow(E) + D_\downarrow(E)}, \quad \text{and} \quad P_\downarrow = 1 - P_\uparrow. \qquad (3)$$

The densities of states for majority-spin and minority-spin carriers, $D_\uparrow(E)$ and $D_\downarrow(E)$, respectively, in Eq. (3) can be obtained from the microscopic models of ferromagnetic metals [32]. The density matrix $\rho = \begin{pmatrix} \rho_{\uparrow\uparrow} & \rho_{\uparrow\downarrow} \\ \rho_{\downarrow\uparrow} & \rho_{\downarrow\downarrow} \end{pmatrix}$ is used to describe the single particle spin state [28,33]. The electron states in the ferromagnetic material can be generated statistically with random numbers to satisfy Eqs. (2) and (3).

In this work, we take into account two injection mechanisms. One is the thermionic emission by which the electrons surmount the Schottky barrier. The other is the tunneling by which the electrons are injected into semiconductor. The low dimension of the semiconductor heterostructure produces an additional selection rule based on the energy quantization in the direction orthogonal to the QW plane. For simplicity, the one subband approximation for transport in the semiconductor structure is adopted [28].

In Fig. 1, the thermionic emission and tunneling processes for electrons are illustrated. We neglect phonon assisted injection mechanisms [34]. Based on the energy conservation, electrons that can afford the thermionic emission need to satisfy

$$E_x > q\phi_B, \quad (4.a)$$

$$E_z = E_{z1}, \quad (4.b)$$

where $x$ is the coordinate in the channel direction, $z$ is in the direction orthogonal to the QW, and $E_{z1}$ is the energy corresponding to the ground state in the QW. The energy states of the particle after the thermionic emission are

$$E'_x = E_x - q\phi_B, \quad (5.a)$$

$$E'_y = E_y, \quad (5.b)$$

$$E'_z = E_{z1}. \quad (5.c)$$

The spin is assumed being conserved during the thermionic emission. That is

$$\rho' = \rho. \quad (6)$$

With Eqs. (5.a)-(5.c), three components of the wave vector in the semiconductor can be obtained as

$$k_x = \sqrt{2m^* E'_x}/\hbar, \quad (7.a)$$

$$k_y = \pm\sqrt{2m^* E'_y}/\hbar, \quad (7.b)$$

$$k_z = \pm k_{z1}. \quad (7.c)$$

We relate the wave vector, **k**, to the particle velocity, **v**, as $\hbar\mathbf{k} = m^*\mathbf{v}$. This definition does not account for the wave vector splitting due to spin-orbit interaction.

In this study, the profile of the Schottky barrier is determined by the self-consistent solution of charge carrier motion and the Poisson equation. With the given barrier profile, the tunneling point $x_{tp}$ is determined randomly from the Fermi-Dirac distribution in the metal and the tunneling probability [30],

$$T_{tp}(E) = \exp(-\frac{2}{\hbar}\int_0^x \sqrt{2m^*[E_c(\xi)-E]}\,d\xi), \qquad (8)$$

based on the WKB method, where $x$ is determined by $E_c(x)=E$. The electron in the ferromagnetic metal that is able to tunnel to the location $x_{tp}$ should satisfy

$$E_x = E_c(x_{tp}), \qquad (9.a)$$
$$E_z = E_{z1}. \qquad (9.b)$$

After the tunneling, the electron has the energy state

$$E'_x = E_x - E_c(x_{tp}) = 0, \qquad (10.a)$$
$$E'_y = E_y, \qquad (10.b)$$
$$E'_z = E_{z1}. \qquad (10.c)$$

In the semiconductor heterostructure, the particle spin state is influenced by the spin-orbit interactions [28]. It is not conserved during the tunneling [35,36]. In the model, the spin density matrix after the tunneling is expressed as

$$\rho' = e^{-i\Phi}\rho e^{i\Phi}, \qquad (11)$$

where parameter $\Phi$ is defined by the spin-orbit coupling (see the next section) as

$$\Phi = [\beta <k_{z1}^2>(\sqrt{E_y/E_x}\sigma_y - \sigma_x) + \eta(\sqrt{E_y/E_x}\sigma_x - \sigma_y)]\frac{x_{tp}m^*}{\hbar^2}, \qquad (12)$$

and $\sigma_x$ and $\sigma_y$ are Pauli matrixes. The wave vector after the tunneling is

$$k_x = 0, \qquad (13.a)$$
$$k_y = \pm\sqrt{2m^*E'_y}/\hbar, \qquad (13.b)$$
$$k_z = \pm k_{z1}. \qquad (13.c)$$

### III. Spin-Polarized Transport in the Quantum Well

The initial states for electrons injected into the semiconductor from the ferromagnetic metal are determined by the model presented in Sec. II. After the injection, electrons start traveling in the QW subjected to scatterings, spin-orbit interactions, applied electric field and the space-charge electric field described by the Poisson equation. The Monte Carlo scheme has been used in several studies of the spin-polarized transport influenced by spin-orbit interaction [28,37,38,39]. We use an ensemble Monte Carlo approach developed previously for spin-polarized transport in low dimensional semiconductor affected by spin-orbit interaction [28]. The model treats the particle dynamics in space classically. The electron spin dynamics is described by the spin density matrix, and the spin evolves during the "free flight" [40] influenced by both Rashba [41] and Dresselhaus [42] effects,

$$\rho(t + \Delta t) = e^{-iH_{SO}\Delta t/\hbar}\rho(t)e^{iH_{SO}\Delta t/\hbar}, \qquad (14)$$

where the spin-orbit interaction is described by

$$H_{SO} = H_R + H_D^1 + H_D^3 \qquad (15)$$

with the Rashba term,

$$H_R = \eta(\sigma_x k_y - \sigma_y k_x), \qquad (16)$$

the linear Dresselhaus term,

$$H_D^1 = \beta\langle k_z^2\rangle(\sigma_y k_y - \sigma_x k_x), \qquad (17)$$

and the third-order correction for the Dresselhaus term,

$$H_D^3 = \beta(\sigma_x k_x k_y^2 - \sigma_y k_y k_x^2). \qquad (18)$$

Eqs. (17) and (18) are written in the crystallographic coordinate system.

## IV. Simulation Results for Fe/GaAs Spin Injection

We have applied the developed model to study the spin injection from a magnetic Fe contact through a Schottky barrier into a single QW of an asymmetrically doped $Al_{0.4}Ga_{0.6}As/GaAs/Al_{0.4}Ga_{0.6}As$ heterostructure with a GaAs layer width, $w = 10$ nm. The bcc Fe lattice constant is about twice as small as the GaAs one [43]. Therefore, the Fe lattice matches well with the GaAs crystal structure in the orientation (0,0,1)Fe/ (0,0,1)GaAs. For such a configuration, the Schottky barrier interface with low material interdiffusion and small concentration of defects have been realized by the electrodeposition technique [44] and molecular beam epitaxy [45]. The Schottky barrier height in the simulation is $q\phi_B = 0.72$ eV [46]. The spin densities of states for up and down spins in the magnetized ferromagnetic material Fe close to the interface are taken from the reference [32]. The QW depth is $\Delta E_c = 0.32$ eV [47]. The coupling constants for spin-orbit interaction in 2DEG are usually measured indirectly [12,13,48]. This results in large variation of values for the extracted parameters. Moreover, it is difficult to distinguish Rashba and Dresselhaus interactions in most cases. We use the following spin-orbit coupling constants: $\beta = 28$ eV·Å$^3$ [49] for the Dresselhaus effect [42] and $\eta = 0.005$ eV·Å for the Rashba effect [41]. The latter parameter is chosen to be compatible with the recent experimental measurements [13]. In this case, the Dresselhaus term is dominant for the spin-obit interaction.

In the simulation, the channel length for spin-polarized transport is $l = 0.7$ μm. The injection takes place at the Fe/GaAs (source) interface at $x = 0$. Because we are investigating the spin-polarized injection through the Schottky barrier and the subsequent transport in the QW, the collection (drain) is assumed non-selective and made of heavily n-doped GaAs. In the Monte Carlo simulation, the time step is chosen as $\Delta t = 1$ fs, and the temperature is 300 K. A drain-source voltage $V_{DS}$ equal to 0.1 V is applied to reverse-bias the Schottky contact, which is in favor of spin injection from the ferromagnetic metal Fe into the semiconductor GaAs.

The calculated energy band diagram of the device, displayed in Fig. 1, is determined by the self-consistent solution of electron motion and the Poisson equation. The large bending near the contact (0 < $x$ < 0.08 μm) indicates the existence of a depletion layer where electron concentration is negligible. The probability distribution of spin-polarized electrons injected from the ferromagnetic metal as a function of energy is given in the inset of Fig. 1. The energy of the injected electrons is mainly distributed in the range of 0.54 ~ 0.9 eV which is far above the Fermi surface in the metal. These results indicate that the injection is primarily derived from the tunneling for the given barrier height and the bias condition.

Our simulation results show that the total spin polarization, the ensemble average of the spin vector over all the particles at a given position, reduces to nearly zero in a distance of 20 nm. This is because the significantly large population of non-spin polarized electrons in the device channel outside the "completely depleted" region substantially diminishes the average value. This phenomenon was also pointed out in Ref. [24].

To investigate the spin dynamics of the injected electrons only, we present the simulation results in Fig. 2 for evolution of the spin-polarization in space, excluding effect of the non-polarized electrons, with various injection mechanisms and different orders of spin-orbit couplings. In Figs. 2(a) and 2(d) where only the linear spin-orbit interaction is included, the spin polarization vector rotates about the $x$ axis due to the dominant Dresselhaus spin-orbit interaction term. The Rashba term results in the rotation about the $y$ axis [28].

The spin dephasing observed in all the six cases is due to the electron momentum scattering events [50]. The very strong initial drop of the spin polarization is believed to result from very high electric field, together with an extremely large field gradient, in the depletion region (see Fig. 1). The high field near $x = 0$ accelerates the electrons swiftly, and quick disappearance of the field leads to rapid randomization of electron motion due to scattering. The large random velocity (thus high electron temperature) results in fast dephasing of the spin polarization [40].

Comparison among Figs. 2(a), 2(b), 2(d) and 2(e) indicates that the higher-order Dresselhaus term has a strong influence on the spin dynamics. Similar phenomenon has been studied for spin-polarized transport in a quasi-1D structure [29]. The injected carrier polarizations, determined by the spin-state probability of electrons in the ferromagnetic contact given in Eq. (3) described by the split densities of states, are mostly minority spins. It should be pointed out that transport across the metal/semiconductor interface is also spin-dependent [26]. Inclusion of this spin-dependent process may have a strong influence on the results. However, different models for the spin-state probability and total injected current could be implemented within the frame of the Monte Carlo model presented in Sec. II without much difficulty.

To describe the spin property of the electron gas, several types of characteristics can be introduced [2]. In the considered case, measurement of the spin polarization may not be appreciated due to the large population of non-polarized electrons in the device channel which substantially weakens effect of the spin polarization. On the other hand, spin flux (or spin current) $J_{s\alpha}(x)$ naturally separates the injected spin polarized electrons from the non-polarized background electrons, where $J_{s\alpha}(x)$ is defined as the accumulation of $\hbar k_x S_\alpha / m^*$ per unit volume for all the electrons at the position $x$. In our model, electron-electron scattering is not included, and non-polarized electrons have no effect on the spin flux. Measurement of spin flux is not obvious. However, it can be observed indirectly using the electron spin accumulation in a detector [17,18].

Fig. 3 displays the spin flux corresponding to the cases presented in Fig. 2. It can be seen that each component of the spin flux decays much more slowly than that of the spin polarization. The nonlinear spin-orbit term given in Eq. (18) has the same effect on the spin flux and the spin polarization. To quantitatively demonstrate the decay of spin flux in a similar way to the spin polarization, we define spin current polarization as

$$P^J = \sqrt{\sum_{\alpha=x,y,z} J_{s\alpha}^2} \Big/ J. \qquad (19)$$

Eq. (19) is applicable for spin polarized current rather than pure spin current [51]. It should be noted that the spin current polarization defined in Eq. (19) differs from the particle spin polarization [28,37] accumulated for injected particles only. In the homogeneous case, a simple relation between two characteristics can be obtained [52]. However, in realistic devices where carrier concentration in the channel is highly inhomogeneous, the relation between particle and spin current polarizations is more complicated.

Fig. 4 shows the spin current polarization associated with the same cases in Fig 2. Fig. 4 (a) presents the spin current polarization with the $S_x$ injection, and Fig. 4(b) the $S_y$ injection. The spin flux (or current) dephasing length is found to be greater than the channel length ($l_s > 0.7$ μm) for the injection of either the $S_x$ or $S_y$ spin polarization with only the linear spin-orbit terms, Eqs. (16) and (17). However, with the higher-order term included, the spin flux dephasing length is reduced to the order of 0.1 μm for the injection with the $S_x$ polarization. On the other hand, the $S_y$ injection including the higher-order spin-orbit coupling leads to a dephasing length that is still greater than the channel length.

## V. Discussions

The developed model is based on several approximations, including the fixed Schottky barrier height, the one subband approximation for the QW, and the spin polarization defined by the spin densities of states in the ferromagnetic contact, etc. However, the model presented in this study is able to reflect main effects of the Schottky barrier on the spin polarized transport in spintronic devices utilizing the mechanism of spin injection. The model can also be extended to 3D transport in bulk GaAs to study Spin-LED structure [14,18]. We argue that spin dynamics of the injected current is mostly determined by the distribution of electron momentum **k** through the spin-orbit interaction mechanism. Macroscopically, it can be characterized by drift and thermal energies of the electrons [40,53]. It is assumed that the additional spin scattering mechanisms [54-58] will not considerably change the spin dynamics presented in this study. However, they can be incorporated into the model.

In the ferromagnetic Schottky barrier, a large portion of spin-polarized electrons is injected through the interface with high kinetic energy subjected to extremely high electric field (see Fig. 1 at $x \sim 0$ in the semiconductor). Near the contact, transport is highly non-equilibrium. In fact, nearly ballistic motion is observed for the injected electrons close to the contact. This distinguishes the Schottky structure from another promising design to achieve efficient spin-polarized injection, the magnetic-semiconductor/nonmagnetic-semiconductor p-n junction [59,60]. In the latter case, drift-diffusion (i.e. quasi-equilibrium transport) dominates spin injection.

## VI. Conclusions

We have developed a Monte Carlo model to describe the spin injection through the Schottky barrier and spin-polarized transport in a semiconductor QW accounting for the Rashba and Dresselhaus spin orbit coupling terms with the third order term included in the Dresselhaus term. Both mechanisms of thermionic emission and tunneling are included in the spin injection model. This approach can be used for simulation of non-equilibrium spin-dependent transport in spintronic devices with a Schottky contact as the source. We presented the simulation results for the spin injection from a 3D ferromagnetic metal into a single quantum well structure through a Schottky barrier with a height of 0.72 eV. The device is reverse-biased at 0.1 V. The spin flux and spin current polarization are introduced to describe the spin dynamics. It is found that the dephasing length of spin current polarization can be greater than the device length (0.7 μm) in the case with only liner spin-orbit coupling. On the contrary, the higher-order Dresselhaus spin-orbit term produces dramatic effect on spin dynamics. The dephasing length can be reduced to a value as low as 0.1 μm for some particular orientation of spin injection. However, for some spin-polarized injection orientation (for example in the *y* orientation), the dephasing length could be maintained to a value greater than 0.7 μm. It is noted that, in realistic spin FETs, evolution of the spin polarization is actually quite different from that of the spin current polarization. In spin FETs, the spin current polarization may be a more suitable characteristic for analysis of spin polarized transport than the particle spin polarization.

*Acknowledgement* - We are grateful to Prof. V. Privman for helpful discussions. One of the authors, S. Saikin, would also like to thank Prof. W. H. Butler for his useful comments. This work was supported by the National Security Agency and Advanced Research and Development Activity under Army Research Office contract DAAD-19-02-1-0035, and by the National Science Foundation grant DMR-0121146.

**Figure Captions**

Fig. 1  Energy band diagram including the Schottky barrier and the Fermi level in the metal, together with two injection mechanisms (thermionic emission and tunneling). The energy distribution of the injected spin-polarized electrons from the ferromagnetic metal is shown in the inset.

Fig 2.  Spin polarization evolution. The injected electrons are spin-polarized (a)-(c) in the $x$ orientation and (d)-(f) in the $y$ orientation. The injection percentages and the spin-orbit interaction are chosen as follows: (a) and (d) 100% spin-polarized injection with linear spin-orbit interaction; (b) and (e) 100% spin polarized injection with linear and nonlinear spin-orbit interactions; (c) and (f) injection determined by the densities of states given in Eq. (3) with linear and nonlinear spin-orbit interactions.

Fig. 3  Spin flux evolution. The injected electrons are spin-polarized (a)-(c) in the $x$ orientation and (d)-(f) in the $y$ orientation. The injection percentages and the spin-orbit interactions are chosen as follows: (a) and (d) 100% spin-polarized injection with linear spin-orbit interaction; (b) and (e) 100% spin polarized injection with linear and nonlinear spin-orbit interactions; (c) and (f) injection determined by the densities of states given in Eq. (3) with linear and nonlinear spin-orbit interactions.

Fig. 4  Evolution of the spin current polarization. The injected electrons are spin-polarized (a) in the $x$ orientation and (b) in the $y$ orientation. Case I: 100% spin-polarized injection with linear spin-orbit interaction; Case II: 100% spin-polarized injection with linear and non-linear spin-orbit interactions; Case III: spin-polarized injection determined by the densities of states given in Eq. (3) with linear and non-linear spin-orbit interactions.

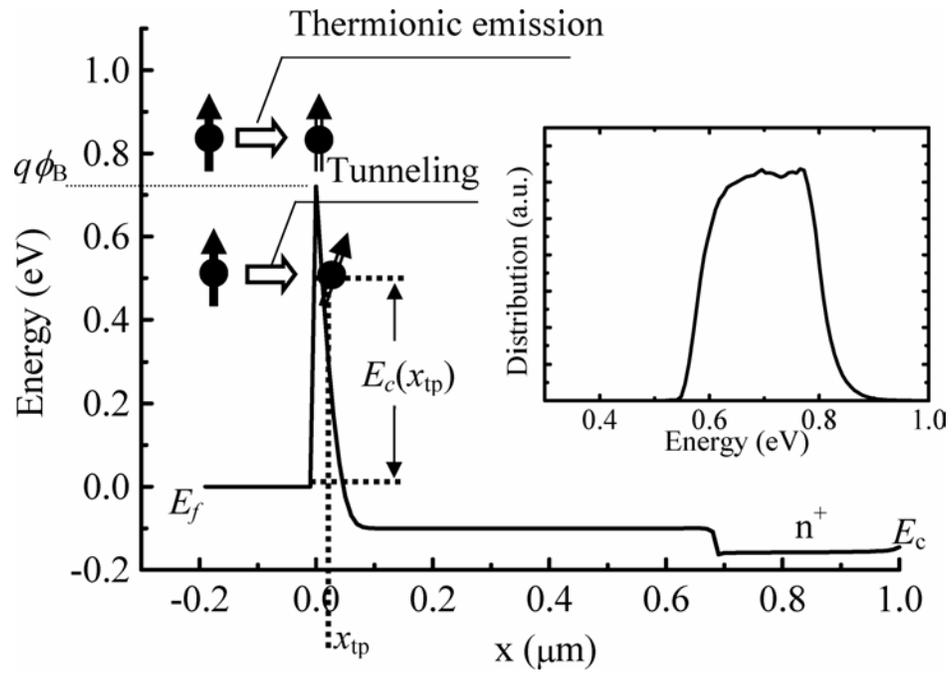

Fig. 1

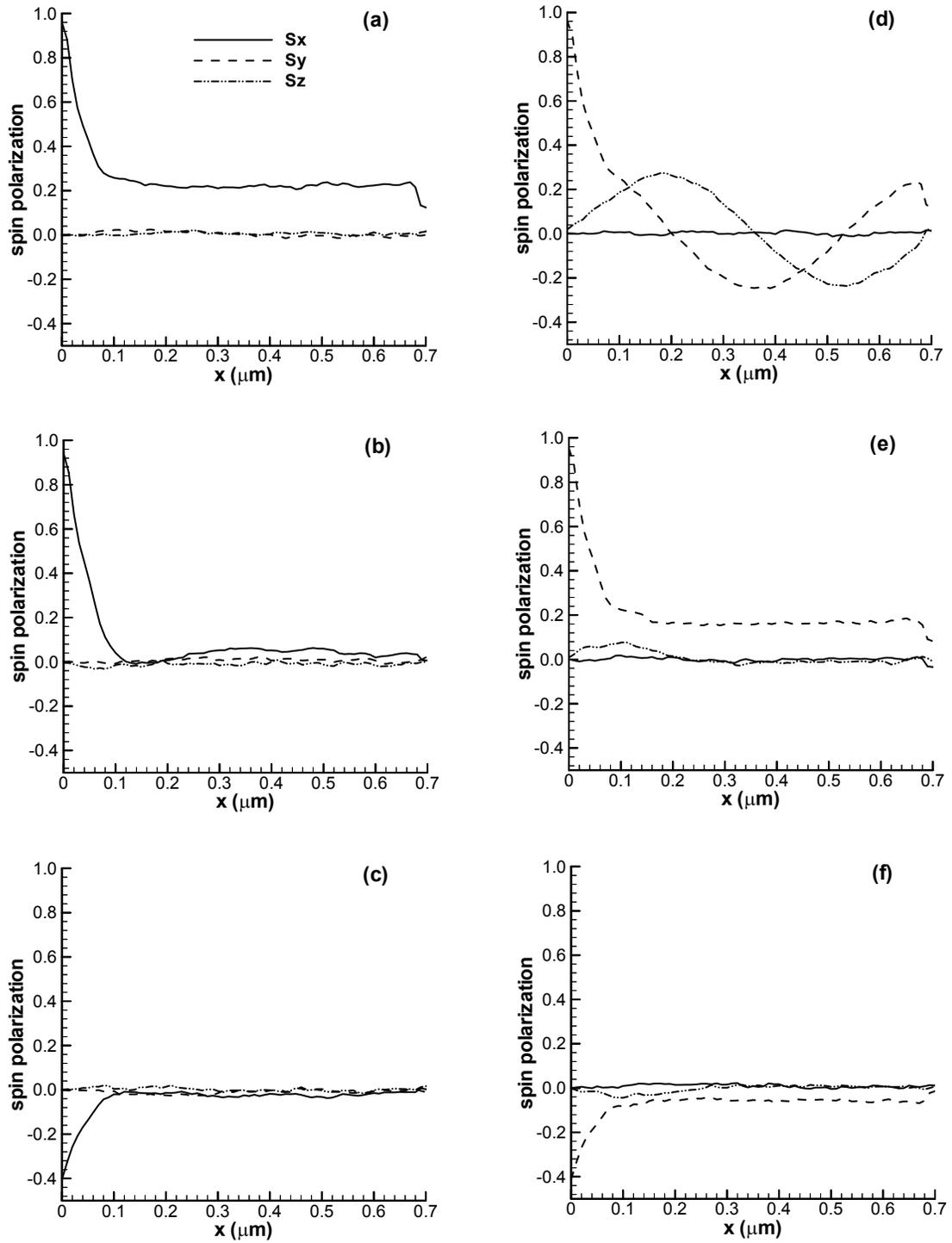

Fig. 2

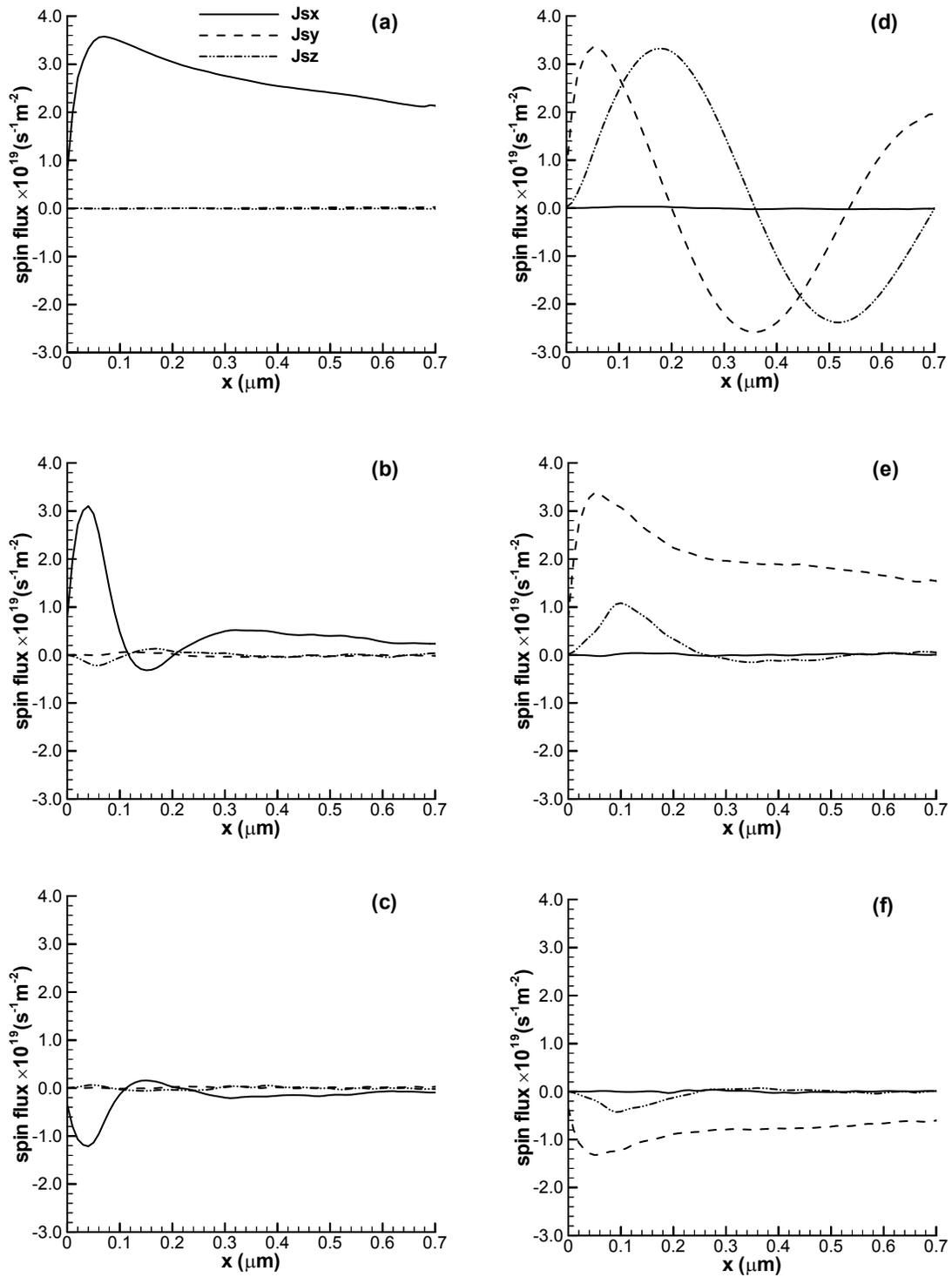

Fig. 3

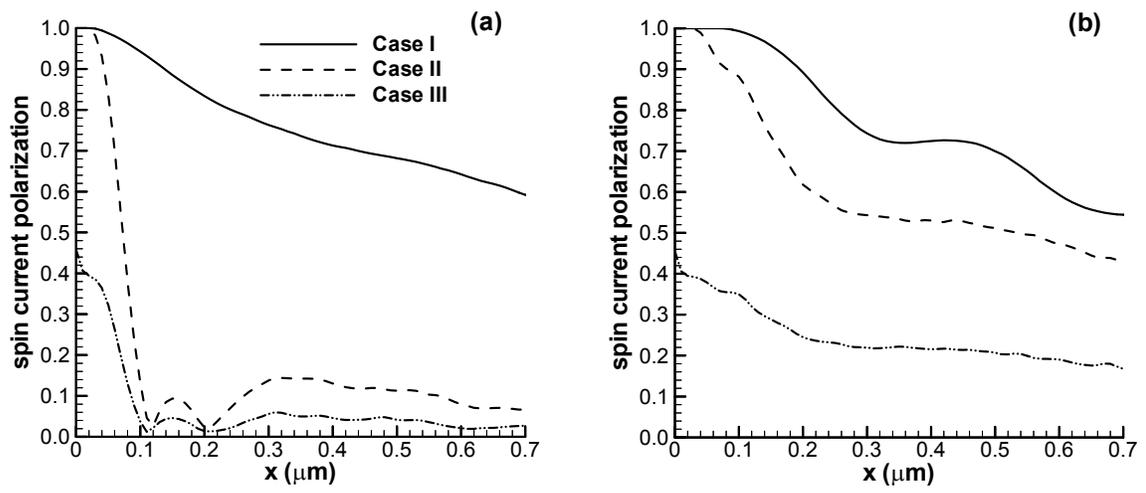

Fig. 4